\documentclass{iaus}

\usepackage{courier}
\usepackage[scaled=0.92]{helvet}
\usepackage{times}
\usepackage{natbib}
\usepackage{graphicx}

\usepackage{color}
\definecolor{myRed}{rgb}{0.84,0.08,0.52}
\definecolor{black}{rgb}{0,0,0}
\definecolor{blue}{rgb}{0,0,1}
\usepackage{hyperref}
\hypersetup{
    final=true,
    pageanchor=true,
    colorlinks=true,
    breaklinks=true,
    linkcolor=blue,
    citecolor=blue,
    urlcolor=blue,
    pdfpagemode=UseNone,
    pdftitle={Velocity fields in and around
        sunspots at the highest resolution},
    pdfauthor={Carsten Denker and Meetu Verma},
    pdfsubject={Solar Physics},
pdfkeywords={Sun: atmospheric motions, Sun: photosphere, Sun: sunspots, Sun:
    magnetic fields}}

\newcommand\arcsec{\mbox{$^{\prime\prime}$}}

\title[Velocity fields in and around sunspots]{Velocity fields in and around
    sunspots\\ at the highest resolution}
\author[C.\ Denker and M.\ Verma]{Carsten Denker and Meetu Verma}
\affiliation{
    Astrophysikalisches Institut Potsdam,
    An der Sternwarte 16,
    D-14482 Potsdam,
    Germany\break
    email: \href{mailto:cdenker@aip.de}{\textsf{cdenker@aip.de}}
        and \href{mailto:mverma@aip.de}{\textsf{mverma@aip.de}}}
\pubyear{2011}
\volume{273}
\pagerange{1--6}
\date{\today}
\jname{Physics of Sun and Star Spots}
\editors{D.P.\ Choudhary \& K.G.\ Strassmeier, eds.}

\begin{document}
\maketitle


\begin{abstract}
The flows in and around sunspots are rich in detail. Starting with the Evershed
flow along low-lying flow channels, which are cospatial with the horizontal
penumbral magnetic fields, Evershed clouds may continue this motion at the
periphery of the sunspot as moving magnetic features in the sunspot moat.
Besides these well-ordered flows, peculiar motions are found in complex
sunspots, where they contribute to the build-up or relaxation of magnetic shear.
In principle, the three-dimensional structure of these velocity fields can be
captured. The line-of-sight component of the velocity vector is accessible with
spectroscopic measurements, whereas local correlation or feature tracking
techniques provide the means to assess horizontal proper motions. The next
generation of ground-based solar telescopes will provide spectropolarimetric
data resolving solar fine structure with sizes below 50~km. Thus, these new
telescopes with advanced post-focus instruments act as a `zoom lens' to study
the intricate surface flows associated with sunspots. Accompanied by
`wide-angle' observations from space, we have now the opportunity to describe
sunspots as a system. This review reports recent findings related to flows in
and around sunpots and highlights the role of advanced instrumentation in the
discovery process.
\end{abstract}
\firstsection


\section{Probing the velocity fields in and around sunspots}

\begin{figure}[t]
\centerline{\includegraphics[width=\textwidth]{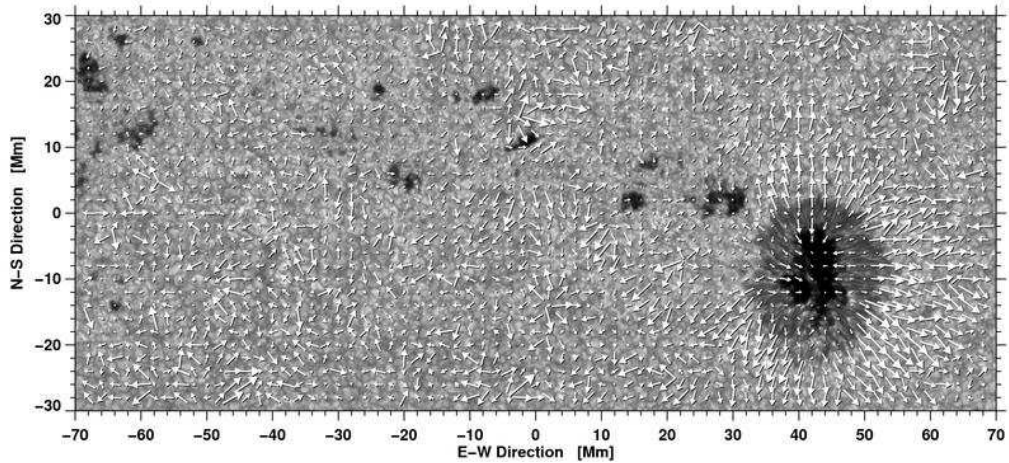}}
\caption{The Japanese \textit{Hinode} mission placed for the first time a
    telescope into space, which can resolve the fine structure of sunspots. SOT
    feeds an instrument suite for high-resolution imaging and
    spectropolarimetry. These novel instruments provide access to the flow
    fields in and around sunspots. Displayed is a G-band image of active region
    NOAA~10921 observed near disk center on 2006 November~3. Direction and
    magnitude of horizontal flows are indicated by vectors. These
    high-resolution data set the stage for this review providing motivation for
    the next generation of ground-based telescopes and instruments.}
\label{FIG01}
\end{figure}

New instruments and observing capabilities have advanced our knowledge about
plasma motions in and around sunspots and their interaction with the magnetic
fields. Therefore, by introducing current and future instruments for
high-resolution studies of the Sun, we will set the stage for this review of the
intricate flow fields of sunspots. The line-of-sight velocity can be derived
from a Doppler-shifted spectral line profile, and its height dependence can be
inferred by carefully selecting lines, which originate at different layers in
the solar atmosphere, or by measuring the bisectors of spectral lines. Local
correlation or feature tracking are the methods of choice in determining
horizontal proper motions (see Fig.~\ref{FIG01}). In principle, spectroscopic
and imaging techniques provide access to the three-dimensional velocity field.

Instruments commonly used to measure solar velocity fields can be placed into
five broad categories ordered according to their spectral resolving power: (1)
imaging with interference filters, (2) imaging with Lyot filters, (3)
line-of-sight velocity and magnetic field measurements using filtergraphs, (4)
imaging spectropolarimeters, and (5) long-slit spectrographs. We introduce
observations obtained with the instruments$^{1,2,3,5}$ (superscripts indicate
the instrument category) of the \textit{Hinode} Solar Optical Telescope
\citep[SOT,][]{Tsuneta2008}, the Interferometric Bidimensional Imaging
Spectrometer \citep[IBIS$^{4}$,][]{Cavallini2006}, the Crisp Imaging
Spectropolarimeter \citep[CRISP$^{4}$,][]{Scharmer2006b}, the G\"ottingen/GREGOR
Fabry-P\'erot Interferometer
\citep[GFPI$^{4}$,][]{BelloGonzalez2008,Denker2010a}, the Tenerife Infrared
Polarimeter \citep[TIP$^{5}$,][]{Collados2007}, and the Polarimetric Littrow
Spectrograph \citep[POLIS$^{5}$,][]{Beck2005}.

The recent success of the Japanese \textit{Hinode} mission \citep{Kosugi2007}
has certainly proven that high resolution observations pave the way for
advancing solar physics. On the ground first observations have been obtained
with the New Solar Telescope \citep[NST,][]{Denker2006b,Goode2010}, while the
GREGOR solar telescope \citep{Volkmer2010a} awaits commissioning upon delivery
of its primary mirror. With the Advanced Technology Solar Telescope
\citep[ATST,][]{Rimmele2010} beginning construction and the  European Solar
Telescope \citep[EST,][]{Collados2010} finishing the design and development
phase, further progress and new discoveries are expected when approaching the
fundamental spatial scales for physical processes on the Sun.

A short review can never be complete but we at least strove to be up-to-date.
Consequently, our approach had to be a very subjective one motivated by the
imminent commissioning of the GREGOR solar telescope and the upcoming science
demonstration time with the GFPI.

\section{The intricate flow fields of simple sunspots}


\subsection{Convective signatures in the umbra}

The existence of convective signatures in the umbra has been debated for many
years. High-resolution observations with CRISP pinpointed strong upflows of up
to 1.4~km~s$^{-1}$ in deep photospheric layers, which are associated with bright
umbral dots \citep{Ortiz2010}. The height dependence of line-of-sight velocities
was determined from the bisectors of coarsely sampled spectral lines. The umbral
dots are identified with columns of hot, rising material with weaker and more
inclined magnetic fields as compared to the umbral background. To establish
overturning convective energy transport in the umbra, downflows have to exist,
which were indeed observed (0.4--1.0~km~s$^{-1}$) in confined areas at the
periphery of umbral dots as predicted by theory. Some elongated umbral dots
exhibit dark lanes in their centers, where the magnetic field is even weaker
and more inclined. These umbral dots resemble `coffee beans' because of this
peculiar intensity pattern, which originates from the accumulation of material
at the top of rising plumes increasing the density and elevating the opacity
\citep{Schuessler2006}. As pointed out by \citet{Ortiz2010}, the velocity
gradient with height in umbral dots imposes an observational challenge,
\textit{i.e.}, the Doppler shifts are hard to observe in the line core --
explaining the scarcity of reported umbral downflows, which are a phenomenon of
the deep photosphere. In addition, small-scale features such as umbral dots
change size and velocity within a few minutes, thus necessitating observations
with high spatial and temporal resolution.

Scrutinizing light-bridges separating different umbral cores,
\citet{RouppevanderVoort2010} found another remarkable feature. Even though
similar in appearance to granulation, light-bridges have dark central lanes with
a width of about 400~km, where upflows reside, which are strongest in deep
photospheric layers. Typical velocities are around 0.5~km~s$^{-1}$ but reach as
high as 1.0~km~s$^{-1}$. However, the direction of the flows is just opposite to
the downflows in intergranular lanes. Similar to the dark lanes in umbral dots,
the narrow, dark channels of light-bridges have weaker and more inclined
magnetic fields.


\subsection{Penumbral fine structure and Evershed flow}

In the penumbra, the Evershed flow denotes a radial outflow of gas, which is
aligned with more horizontally placed magnetic flux tubes in dark penumbral
filaments. Theoretical models based on the thin flux tube approximation arrive
at steady as well as time-dependent flows driven by pressure differences. The
`moving flux tube' model \citep[\textit{e.g},][]{Schlichenmaier2002} explains
much of the observed fine structure. However, `steady siphon-flow' models still
remain viable and provide strong arguments against super-Alfv\'enic,
`sea-serpent'-like flows \citep{Thomas2005}, namely that they are
gravitationally unstable.

The average vertical velocity field of the quiet Sun as observed by the
\textit{Hinode} spectropolarimeter \citep{Franz2009} is always dominated by
upflows, whereas the penumbra shows a different behavior. Here, upflows cover a
larger area for Doppler velocities below 0.4~km~s$^{-1}$, in particular in the
inner penumbra. However, at velocities above 0.6~km~s$^{-1}$ downflows have a
larger areal coverage. This leads to a net downflow of more than 0.1~km~s$^{-1}$
for the entire penumbra. The upflows of the inner penumbra are typically
elongated an possess an aspect ratio of about five. In contrast, the largest
downflows in the penumbra of up to 9~km~s$^{-1}$ at the outer penumbral boundary
exceed even the largest quiet Sun values of about 3~km~s$^{-1}$ and have a shape
closer to circular.

Penumbral grains migrate inwards in the inner penumbra. The Evershed flow begins
at the leading edge of the penumbral grains \citep{Ichimoto2007}, which have
been identified with the footpoints of hot upflows in strongly inclined flux
tubes \citep{Rimmele2006}. The Evershed flow then turns horizontal and follows
preferentially the dark cores of the penumbral filaments \citep{Scharmer2002},
where the magnetic field is more horizontal. The striking  dark-cored penumbral
filaments are much easier to discern in polarized light than in continuum images
\citep{BellotRubio2007}. Their spectral line profiles are very asymmetric, which
hints at multiple magnetic field components along the line-of-sight or within
the resolution element. The dark cores have a lateral extend of less than 200~km
and exhibit magnetic fields, which are weaker by 100--150~G as compared to the
lateral brightenings. The Evershed flow reaches velocities up to and even
exceeding the photospheric sound speed of about 7~km~s$^{-1}$. The Evershed flow
shows temporal variations on scales of 10--15~min, which also corresponds to the
intensity variations of Evershed clouds. Small patches of opposite polarity and
strong downflows are observed throughout the outer penumbra indicating that some
penumbral field lines already return to the interior well within the penumbra
itself \citep{SainzDalda2008}. Selecting spectral lines with contribution
functions covering the deep photosphere or bisector analysis are the means to
determine the height dependence of the Evershed flow, which increases in
strength with depth. The Evershed flow is not stationary. Coherent flow patches
(Evershed clouds) can be traced from within the sunspot to move away from the
spot \citep{CabreraSolana2006}, where they can be associated with moving
magnetic features (MMFs).


\subsection{Moat flow and moving magnetic features}

\citet{MartinezPillet2009} report that the Evershed flow continues at least
sporadically outside the penumbra into the sunspot moat, \textit{i.e.}, not all
field lines, which carry the Evershed flow, submerge below the photosphere at
the penumbra's outer boundary. Analyzing spectropolarimetric \textit{Hinode}
data, \citet{Shimizu2008} find high-speed (supersonic) downflows by analyzing
Stokes-$V$ profiles. If such profiles become more complicated, \textit{i.e.}, if
they have multiple lobes, then using the zero-crossing of the Stokes-$V$ profile
is not a good indicator of flow velocities. This argues in favor of instruments
capable of resolving spectral line profiles and not filtergraph systems, which
deliver Dopplergrams and magnetograms prone to erroneous interpretation. The
observed downflows occur in three distinct locations: (1) the outer penumbral
boundary (MMFs), (2) the edge of the umbra in absence of penumbral structures,
and (3) near small-scale field concentrations in the sunspot moat (convective
collapse). These features have in common a pointlike appearance with diameters
of about 1\arcsec. They are transient features with lifetimes from a few minutes
to about 30~min but on average the lifetimes tend towards the lower end of this
range. On the other hand, in long-duration observations, filamentary magnetic
features become visible in the sunspot moat \citep{SainzDalda2005} revealing
that MMFs preferentially move along certain pathways. \citet{Balthasar2010}
present another interesting finding, namely that in the inner moat the flow
velocities are higher in the ultra-violet (170~nm) as compared to visible
(500~nm) continuum, while reversing this relationship in the outer moat.

Active region NOAA~10977 contained a bipolar group of pores, which never
developed a penumbra, even though infrared Ca\,\textsc{ii} $\lambda 854.2$~nm
observations revealed superpenumbral structures (a worthwhile research topic on
its own). In a coordinated observing campaign with \textit{Hinode} and IBIS,
\citet{Zuccarello2009} detected short, radially aligned magnetic structures at
the periphery of the pore, which however did not bear any resemblance to
penumbral filaments. Consequently, in their absence no indications of the
Evershed effect could be detected. Surprisingly, both moat flow and MMFs were
surrounding the pore questioning their close ties to the Evershed flow. The
presence of MMFs was interpreted as twisted magnetic field lines, which were
pealed away from the vertical flux bundle of the pore by the (super)granular
flow.


\subsection{Decay of sunspots}

A mechanism to remove magnetic flux from a sunspot is the interaction of
penumbral filaments at the edge of the sunspot with the granulation, which
erodes the magnetic field of the sunspot \citep{Kubo2008}. Flux detaches in
the form of MMFs from the  more vertical background fields of the uncombed
penumbra, thus contributing to the decay of sunspots. Other types of MMFs, which
are related to the magnetized Evershed flow, do not change the net magnetic
flux. These dark penumbral filaments with strong horizontal magnetic fields
often reach into the moat region. This goes along with a non-stationary
penumbral boundary, which advances and retracts with respect to an average
position.

In the last stage of sunspot decay, only a pore without a penumbra remains
\citep{BellotRubio2008}, where small finger-like, weak, and almost horizontal
magnetic features of opposite polarity can be recognized at the pore's boundary.
They extend up to about 1.5~Mm and have blue-shifted Stokes profiles indicative
of upflows with speeds of 1--2~km~s$^{-1}$. This could be the remnants of
magnetic field lines, which previously carried the Evershed flow. No longer held
down by the mass provided by the Evershed flow, they become boyant and lift of
to vanish in the chromosphere. In general, the question remains open, where
penumbral flux tubes end. At least in the final stages of a decaying sunspot,
alternatives might exist to the notion that they just bend below the surface at
the periphery of the penumbra. This problem is also tied to the matter of mass
continuity in the Evershed flow, since sources and sinks still need to be
unambiguously identified and the balance between inflows and outflows has to be
established. If the mass would be supplied by the flux rope rising through the
convection zone, which initially led to the emergence of the sunspot, then
decoupling from this flux rope would shut off the mass flow causing the sunspot
to decay. It is noteworthy that the divergence line observed in the middle
penumbra survives the decay process \citep{Deng2007} and even the moat flow is
still detectable long after the penumbra has vanished \citep{Zuccarello2009}.


\section{Peculiar flows in the context of eruptive events}

Flares occur near magnetic neutral lines, where strong magnetic field gradients
exist, and where the horizontal component of the magnetic field is strongly
sheared. The height dependence of horizontal proper motions can be derived by
applying local correlation tracking techniques to images obtained in multiple
spectral regions. Near-infrared images at the opacity minimum probe the deepest
photospheric layers, whereas G-band images provide access to higher layers as
compared to continuum images observed in the visible part of the solar spectrum.
\citet{Deng2006} presented such a multi-wavelength study of active region
NOAA~10486 -- one of the most flare-prolific regions of solar cycle 23. Both
horizontal and vertical shear flows with speeds of about 1~km~s$^{-1}$ exist in
the vicinity of magnetic neutral lines. These flows are long-lived and
persistent. Thus, the magnetic field might not be the only agent trigerring
solar flares. The flow speed in shear regions is diminishing with height while
the direction remain essentially the same. Therefore, shear flows are dominant
in the deeper layers of the photosphere. In response to an X10 flare the shear
flows significantly increased, which was interpreted as shear release in the
overlying magnetic fields or as the emergence of twisted and sheared flux
infusing energy from subphotospheric layers. Shear flows are not just limited to
horizontal flow fields. In a spectropolarimetric study of a B7.8 flare in active
region NOAA~10904, \citet{Hirzberger2009} detected (supersonic) downflows of up
to 7~km~s$^{-1}$ in the penumbra. Twisting and interlaced penumbral filaments
are no longer radially aligned with respect to the major sunspot and become
almost tangential. However, the Evershed flow remains aligned with the magnetic
field. Islands of opposite polarity and complex flows require a
three-dimensional topology necessitating high spectral resolution to capture
these features in multi-lobed Stokes-$V$ profiles.

Twisted and sheared penumbral filaments, photospheric shuffling of footpoints,
and rapid motion as well as rotation of sunspots can all destabilize the
magnetic fields above an active region. In a multi-wavelength study of active
region NOAA~10960, \citet{Kumar2010} find evidence that helical twist is
accumulated before the flare and then activated to release the stored energy
during the flare. The orientation of penumbral filaments can strongly deviate
from the radial direction in satellite sunspots and $\delta$-configurations.
Photospheric signatures of the flare are the now well-established rapid
penumbral decay and umbral enhancement \citep{Liu2005}, which are
indicative of a rearranged magnetic field topology. In addition to these signatures,
\citet{Gosain2009} noticed the lateral displacements of penumbral filaments
during an X3.4 flare in active region NOAA~10930 on 2006 December~13. This
occurs immediately before (4~min) the flare initiation, when penumbral filaments
move laterally towards the magnetic neutral line, while changing direction and
moving away from it for about 40~min after the flare. The energy involved in the
lateral displacements is only a few percent of the total energy released in the
flare.


\section{Conclusions}

Space observations are not only free from seeing but, depending on orbit, do not
have to cope with the day night-cycle on Earth. Thus, in principle, high cadence
observations with long coverage become possible to follow magnetic structures of
active regions while they evolve. Such observations from space are only limited
by on-board processing and telemetry. Despite the availability of long
time-series with consitstent quality only few systematic and comprehensive
studies of flow fields are available. The majority of investigations have been
limited to case studies. This is an opportunity for the time yet to come to
explore this enormous database with the aim to put the still fragmented puzzle
of flows in and around sunspots together into a comprehensive picture. In this
respect, high-cadence vector magnetograms of the recently launched Solar
Dynamics Observatory (SDO) will certainly advance our knowledge regarding
eruptive events on the Sun.

The new and upcoming instrumental capabilities on the ground will drastically
improve angular resolution. Adaptive optics and future multi-conjugate adaptive
optics will allow to capture time-series of spectropolarimetric data, which are
longer than the lifetime of solar (small-scale) structures. The polarimetric
sensitivity and spectral resolution of these instruments including infrared
capabilities will produce multi-dimensional data sets suitable for advanced
spectral inversion techniques. For the first time, physical quantities become
accessible at the fundamental scales of physical processes on the Sun.


\begin{acknowledgments}
MV expresses her gratitude for the generous financial support by the German
Academic Exchange Service (DAAD) in the form of a PhD scholarship. CD
acknowledges a DAAD travel grant facilitating his attendance at the IAU
Symposium.
\end{acknowledgments}



\end{document}